\documentclass{article}

\usepackage{natbib}
\usepackage{amsmath}

\usepackage[preprint]{neurips_2024_ml4ps}

\usepackage[utf8]{inputenc} 
\usepackage[T1]{fontenc}    
\usepackage{hyperref}       
\usepackage{url}            
\usepackage{booktabs}       
\usepackage{amsfonts}       
\usepackage{nicefrac}       
\usepackage{microtype}      
\usepackage{xcolor}         
\usepackage[pdftex]{graphicx}
\usepackage{ulem} %

\title{Probabilistic Galaxy Field Generation with Diffusion Models}

\author{%
  Tanner Sether \\
  Michigan Technological University\\
  Houghton, MI 49931 \\
  \texttt{tmsether@mtu.edu} \\
  \And
  Elena Giusarma \\
  Michigan Technological University \\
  Houghton, MI 49931 \\
  \texttt{egiusarm@mtu.edu} \\
  \AND
  Mauricio Reyes-Hurtado \\
  Michigan Technological University \\
  Houghton, MI 49931 \\
  \texttt{mrhurtad@mtu.edu} \\
}

\begin{document}

\maketitle

\begin{abstract}

In the era of precision cosmology, the ability to generate accurate and large-scale galaxy catalogs is crucial for advancing our understanding of the universe. With the flood of cosmological data from current and upcoming missions, generating theoretical predictions to compare with these observations is essential for constraining key cosmological parameters. While traditional methods, such as the Halo-Occupation Distribution (HOD), have provided foundational insights, they struggle to balance the need for both accuracy and computational efficiency. High-fidelity hydrodynamic simulations offer improved precision but are computationally expensive and resource-intensive.
In this work, we introduce a novel machine learning approach that harnesses Convolutional Neural Networks (CNNs) and Diffusion Models, trained on the CAMELS simulation suite, to bridge the gap between computationally inexpensive dark matter simulations and the galaxy distributions of more costly hydrodynamic simulations. Our method not only outperforms traditional HOD techniques in accuracy but also significantly accelerates the simulation process, offering a scalable solution for next-generation cosmological surveys. This advancement has the potential to revolutionize galaxy catalog generation, enabling more precise, data-driven cosmological analyses.

\end{abstract}
\section{Introduction}
\vspace{-0.1 in}
Understanding the formation and evolution of galaxies is a fundamental goal in astrophysics and cosmology. Upcoming large-scale surveys (LSS), such as Euclid~\citep {Euclid} and LSST~\citep{lsst}, will provide unprecedented data, offering new opportunities to constrain key astrophysical and cosmological parameters. Extracting maximum information from these surveys requires comparisons with simulations, particularly hydrodynamic simulations, which are computationally expensive and limit the exploration of parameter space. Simplified models like the Halo-Occupation Distribution (HOD)~\citep{hod} improve efficiency but sacrifice accuracy. Recent advances in machine learning, particularly convolutional neural networks (CNNs), have demonstrated superior performance in both speed and precision over HOD; see for example~\citep{camels}~\citep{xinyue}.
 \par
In this work, we propose a novel deep learning framework that leverages variational diffusion models (VDMs)~\citep{kingma2023}~\citep{ono} to map dark matter fields from N-body simulations to galaxy fields derived from high-fidelity hydrodynamic simulations. By combining the robustness of CNNs with the probabilistic power of VDMs, our approach surpasses traditional CNN-based methods, providing a scalable and accurate solution for next-generation cosmological analyses.

\section{Methods}
\subsection{Data}

In this work, we use data from the CAMELS project~\citep{camels}, which includes 5,324 hydrodynamic simulations and 5,097 corresponding N-body simulations. These simulations share consistent cosmological and initial conditions and were generated using different subgrid models, such as SIMBA~\citep{simba} and IllustrisTNG~\citep{illustris}.
Each simulation evolves $256^3$ cold dark matter (CDM) particles, with hydrodynamic simulations also evolving $256^3$ gas particles. The evolution spans from redshift $z = 127$ to $z = 0$ in a periodic co-moving volume of $(25 , \text{Mpc}/h)^3$. Subhalos are identified using the SUBFIND algorithm~\citep{subfind}, and galaxies are defined as subhalos with a stellar mass greater than zero~\citep{camels}.
To enhance model performance, 15 2D maps were generated from each simulation by slicing the 3D outputs into five layers along each dimension and averaging over the thickness of each slice. This approach ensures a robust representation of the galaxy distribution.

The training, validation, and testing data for our model are derived from the IllustrisTNG suite of CAMELS. The dataset includes simulations from the Latin Hypercube set, which consists of 1,000 simulations with six varied cosmological and astrophysical parameters. Additionally, we use the Cosmic Variance set, which contains 27 simulations where the parameters remain fixed, but the initial seeds are varied to capture stochastic effects. Example 2D maps of the target fields and corresponding model predictions are shown in Figure~\ref{fig:statmaps}.

\subsection{CNN Model Architecture}
\label{CNN}
The input data for the model consists of the dark matter density fields from N-body simulations, while the target data is the galaxy density fields from hydrodynamic simulations, both evaluated at $z = 0$. Due to the sparsity of galaxy distributions compared to dark matter, accurately capturing the small-scale dynamics of galaxy formation presents a significant challenge. These dynamics are critical, as they encode the complex physical processes driving galaxy clustering.
To address this, we adopt a two-phase architecture following~\citep{xinyue}. In the first phase, a neural network is trained as a binary classifier to predict the probability of galaxy presence in each voxel. In the second phase, the model is optimized further by focusing on voxels with a high probability of containing galaxies, refining the galaxy density predictions.
 \par
This two-phase architecture is flexible, supporting various network configurations for both the classification and regression phases. In the classification phase, the primary objective is to maximize recall while maintaining high accuracy, which is essential for handling the sparsity of the target data. For the regression phase, model performance is evaluated using multiple metrics. Mean squared error (MSE) directly quantifies the accuracy of predictions, while comparisons of the power spectrum assess how well the model reproduces the statistical properties of large-scale structures, particularly on small scales where galaxy clustering is most sensitive.
For the classification phase, we experimented with several architectures, including Inception, UNet, and R2UNet~\citep{inception}~\citep{unet}~\citep{r2unet}. In the regression phase, R2UNet was compared to a Variational Diffusion Model (VDM)~\citep{kingma2023}, highlighting the differences between deterministic (CNN) and probabilistic (VDM) approaches to galaxy field reconstruction, see Section~\ref{results}.

\subsection{Variational Diffusion Model}

\begin{figure}
    \centering
    \includegraphics[width=0.8\linewidth]{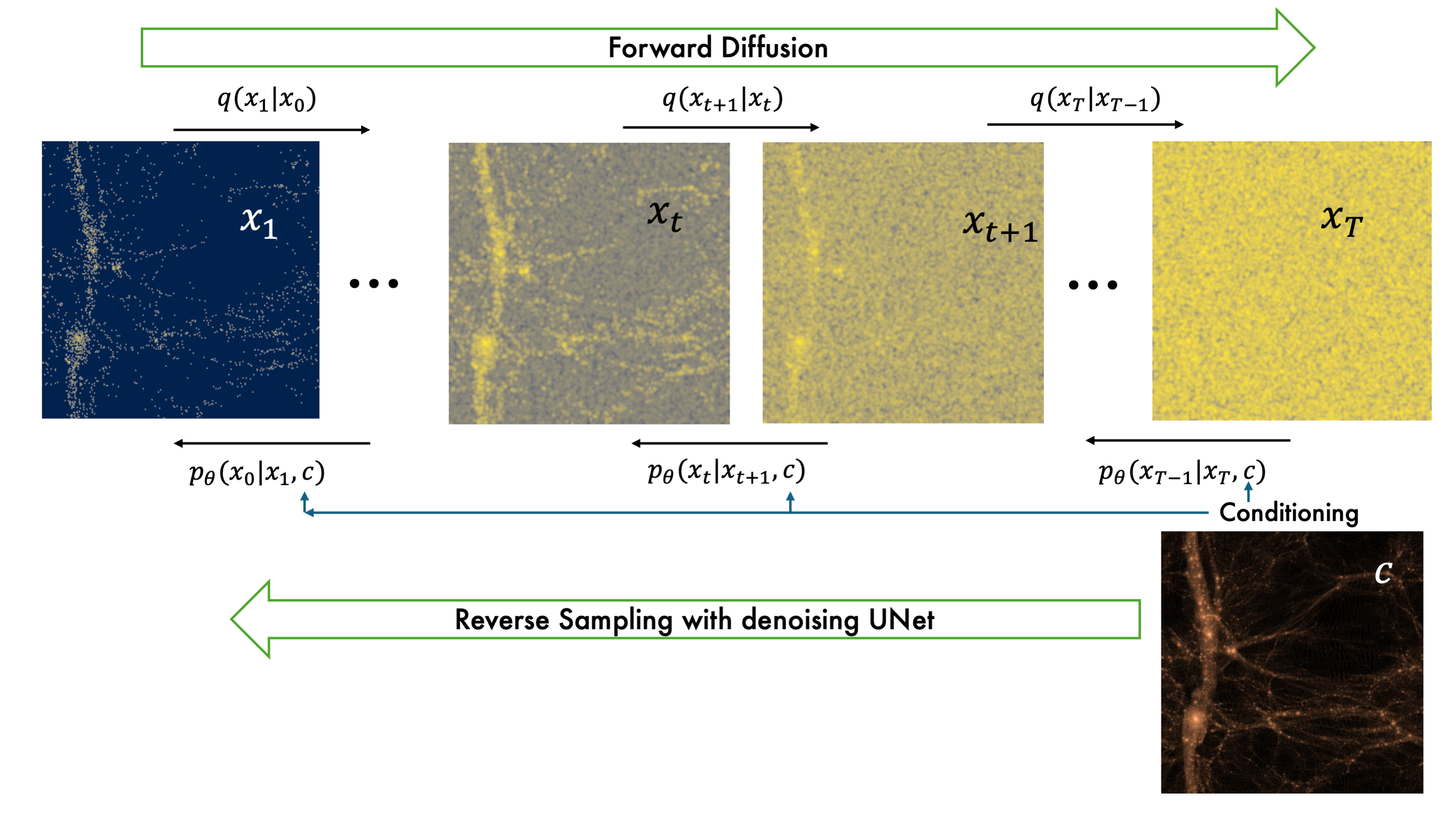}
    \caption{Illustration of the diffusion process employed in the Variational Diffusion Model. The conditional noise schedule is denoted by $q$, and the learned conditional probability estimated by the UNet is represented as $p_\theta$. Figure adapted and modified from \citet{ono}.} 
    \label{fig:VDM}
\end{figure}

The Variational Diffusion Model (VDM) reconstructs galaxy distributions from dark matter fields within a probabilistic framework. The model operates by progressively adding noise to the galaxy fields during a forward diffusion process and then learning to reverse this process step-by-step using a UNet architecture, as illustrated schematically in Figure~\ref{fig:VDM}.

The forward diffusion process systematically introduces noise to the galaxy field, progressively degrading its structure in a controlled and predictable way. The reverse diffusion process, conditioned on the dark matter field, reconstructs the original galaxy field by modeling a sequence of conditional distributions. At each step, the UNet estimates the conditional distribution of the noiseless galaxy field given its noisy counterpart and the dark matter field as input. This probabilistic framework allows the model to approximate the posterior distribution of galaxy fields conditioned on dark matter inputs.
Training the VDM involves minimizing a variational bound on the likelihood of the data. Through this optimization, the UNet learns the parameters required to reverse the diffusion process and accurately denoise the galaxy field. This framework not only provides a robust reconstruction of galaxy distributions but also facilitates uncertainty quantification by modeling the variability inherent in galaxy formation.

\subsection{Benchmark Models}

To evaluate the performance of our model, we compared it against two established methods. The first benchmark is a Halo Occupation Distribution (HOD) model, a widely used parameter-based approach for populating dark matter halos with galaxies~\citep{hod}. The HOD relies on three free parameters: $M_{min}$; the minimum mass for a halo to contain a galaxy, $M_1$; the mass of halos that contain one galaxy on average, and $\alpha$; the power law index. Only halos with masses greater than $M_{min}$ host a central galaxy placed at the halo center. The number of satellite galaxies follows a Poisson distribution with a mean of $(M/M_1)^\alpha$, and these satellites are distributed randomly within the dark matter halo. Given a specific set of $M_1$ and $\alpha$, we optimize $M_{min}$, then fine-tune $M_1$ and $\alpha$ to minimize the mean squared error (MSE) on the power spectrum. The HOD model is utilized as a benchmark because it represents the standard classical approach for efficiently populating dark matter simulations with galaxies. However, this model does not include assembly bias, as we rely on CDM-only snapshots at $z = 0$, making it a simplified but practical comparison. 

The second benchmark is the CNN model, which employs the two-phase architecture described in Section~\ref{CNN}. The CNN captures multiscale information to generate accurate galaxy distributions and serves as a strong deep-learning-based comparison. However, it operates deterministically, unlike our Variational Diffusion Model (VDM), which leverages a probabilistic framework to model the posterior distribution of galaxy fields. This distinction allows the VDM to capture variability and provide uncertainty quantification, setting it apart from traditional benchmarks.

\section{Results}
\label{results}
In this section, we evaluate the performance of the models across both the classification and regression tasks and provide a detailed analysis of the Variational Diffusion Model (VDM). The Inception network demonstrated superior performance in the classification phase, achieving the highest recall, as shown in Table 1. This highlights its capability to handle the sparsity of galaxy distributions effectively, a critical aspect of the classification task. Conversely, while the R2UNet achieved slightly higher accuracy, it underperformed in terms of recall, indicating potential challenges in identifying all galaxy-present regions. The VDM excelled in the regression phase, achieving the lowest mean squared error (MSE) among all tested models. Specifically, it demonstrated a 60\% reduction in MSE compared to the HOD model and a 20\% improvement over the CNN. Notably, the VDM performed exceptionally well at small scales, capturing fine-grained, nonlinear features of galaxy formation more accurately than its benchmarks.

\begin{figure}
    \centering
    \includegraphics[width=0.85\linewidth]{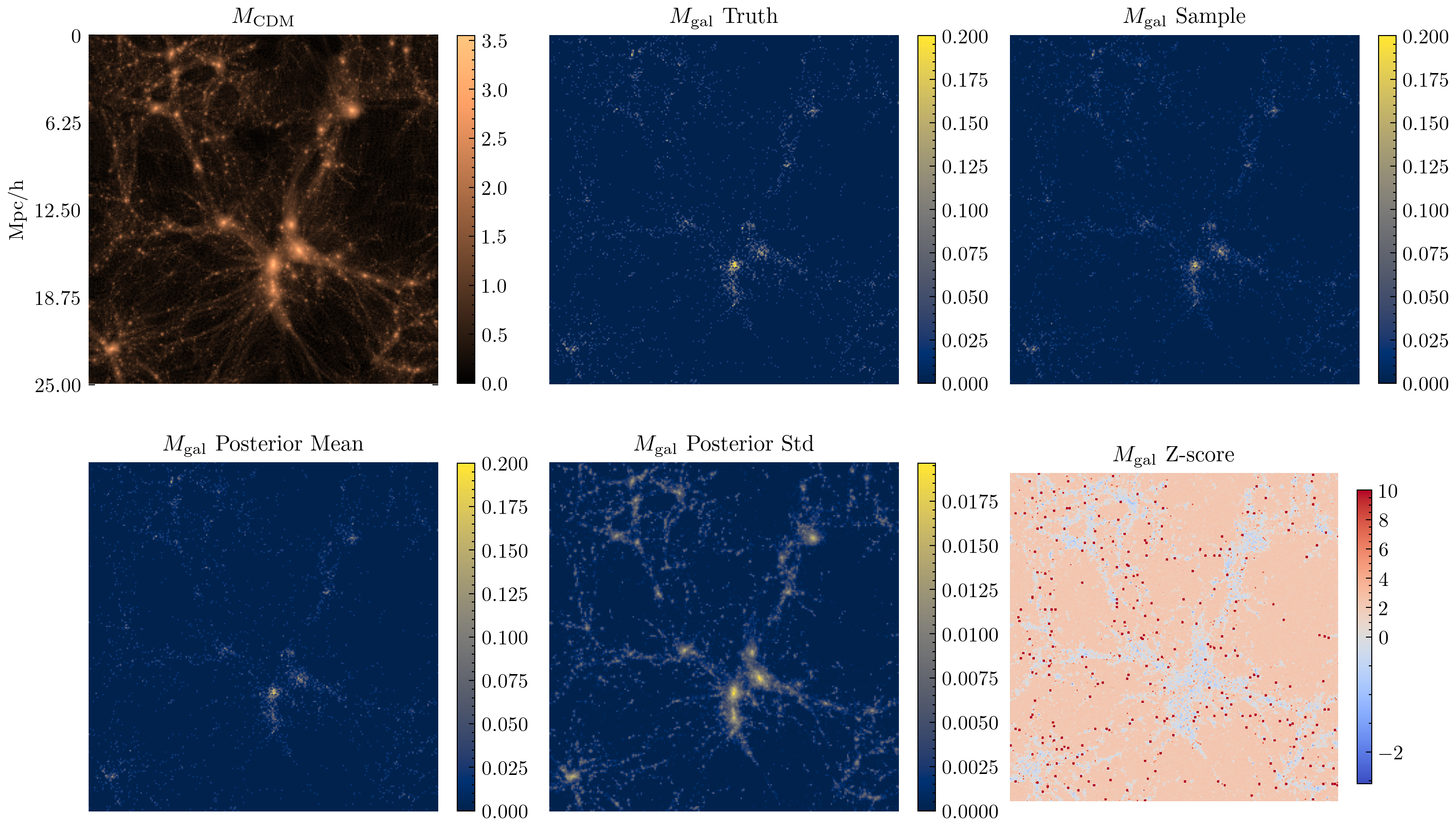}
    \caption{Figure displays six maps from the model: the input CDM field (top left), the true galaxy field (top middle), a single VDM-generated field (top right), the posterior mean (bottom left), the posterior standard deviation (bottom middle), and the Z-score map (bottom right).}
    \label{fig:statmaps}
\end{figure}

\begin{figure}
    \centering
    \includegraphics[width=0.4\linewidth]{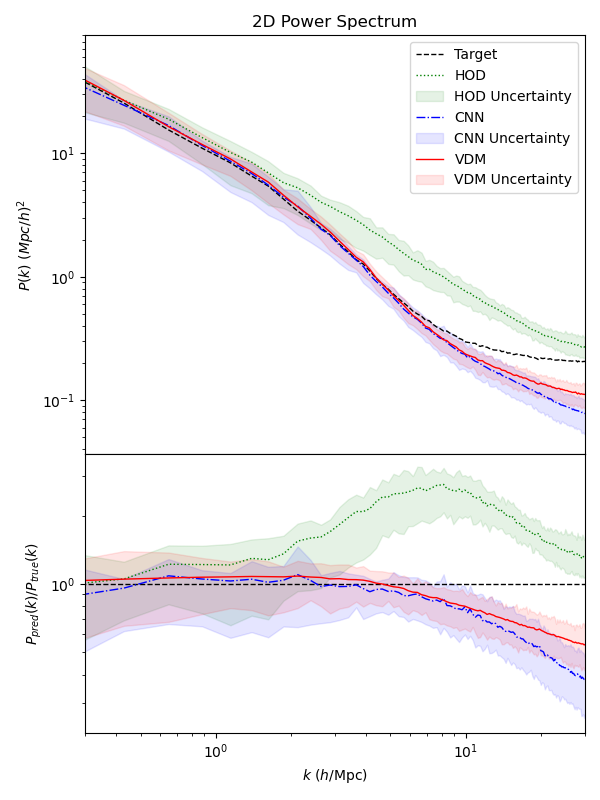}
    \includegraphics[width=0.4\linewidth]{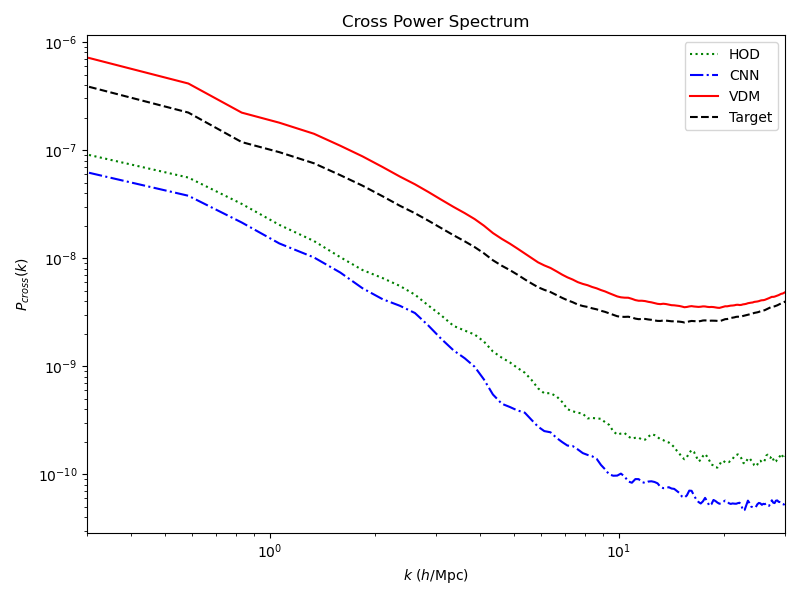}
    \caption{The left plot shows the galaxy power spectra (top) and residuals (bottom) for the target (black line) and the outputs of the models used in this work. Uncertainties are represented by the 25th and 75th percentiles across the test set. The VDM (red line) outperforms the CNN (blue line) and HOD (green line), particularly at small scales, where differences in galaxy formation are most significant. The right plot shows the cross-power spectra, comparing the models to the true distribution.}
    \label{fig:powerplots}
\end{figure}

Figure~\ref{fig:statmaps} provides a visual representation of the VDM’s outputs, including the input CDM map, the true galaxy field, a single VDM-generated sample, and posterior statistics derived from 100 samples The mean map closely aligns with the true galaxy field, with minimal visual discrepancies. The standard deviation map indicates small uncertainty in most regions, with higher uncertainty in sparse areas, reflecting the model's challenges in low-density regions. The Z-score map quantifies these biases: red regions (positive Z-scores) in voids denote overpredicted densities, while blue regions (negative Z-scores) in massive halos indicate underpredicted galaxy concentrations. Grey regions show no significant deviation from the true distribution.
These patterns are consistent with the power spectrum results, where the VDM slightly overestimates galaxy density at large scales and underestimates it at small scales (see Figure~\ref{fig:powerplots}). 

\begin{table}
\centering
\caption{Results from the classification phase of training.}
\label{table:classification}
\begin{tabular}{l l l l l}
\toprule
\textbf{Model} & \textbf{Accuracy} & \textbf{Recall} & \textbf{Precision} & \textbf{Training Time} \\
\midrule
Inception & 97.34\% & 95.38\% & 3.537\% & 6 min \\
R2Unet & 98.24\% & 94.16\% & 5.196\% & 90 min \\
\bottomrule
\end{tabular}
\end{table}

The left plot of Figure~\ref{fig:powerplots} compares the power spectra (top) and the transfer function (bottom) of the VDM model (blue line) with the true galaxy field (black line), the HOD model (green line), and the CNN (red line). The VDM outperforms the HOD and CNN, particularly at large scale ($k < 5 \, h/\text{Mpc}$) and intermediate-scale ($5 \, h/\text{Mpc} < k < 10 \, h/\text{Mpc}$), replicating clustering patterns with improved accuracy. At $k =20 \, h/\text{Mpc}$, the VDM achieves a significant improvement, outperforming the HOD by approximately 50\% and the CNN by 30\%. However, the VDM underperforms at smaller scales ($k > 10 \, h/\text{Mpc}$), reflecting the challenges in modeling nonlinear galaxy formation dynamics in dense environments. This trend aligns with observations from the Z-score map in Figure~\ref{fig:statmaps}.

The right plot of Figure~\ref{fig:powerplots}) highlights the cross-power spectrum, showing the correlation between the predicted galaxy density field and the true field across scales. The VDM consistently exhibits higher cross-power values than the true field, indicating systematic overcorrelation. At large scales ($k < 5 \, h/\text{Mpc}$), the model captures general clustering trends but amplifies the galaxy-dark matter relationship, potentially overestimating their connection. At intermediate scales ($5 \, h/\text{Mpc} < k < 10 \, h/\text{Mpc}$), the overcorrelation persists but moderates slightly. At small scales ($k > 10 \, h/\text{Mpc}$), the cross-power spectrum declines, reflecting challenges in modeling nonlinear dynamics, such as baryonic feedback and dense environmental interactions.
While the VDM effectively captures large and intermediate scale clustering, its overcorrelation across all scales highlights a limitation in its modeling of galaxy-halo connections and the absence of explicit baryonic effects. Future work could address these issues by incorporating additional physical priors or expanding the training dataset with simulations that include baryonic processes.

The VDM requires 3 to 5 times longer training compared to a CNN but can generate 100 samples in approximately 3 minutes, vastly outperforming the HOD model and making it a practical alternative to computationally intensive hydrodynamic simulations~\citep{camels}.

\section{Conclusions}
Our results demonstrate the VDM’s significant advancements over traditional HOD and CNN benchmarks, particularly in capturing large and intermediate-scale galaxy clustering with improved accuracy. While challenges remain in modeling small-scale nonlinear dynamics, the probabilistic framework of VDM provides a robust tool for uncertainty quantification and Bayesian parameter inference. By modeling the posterior distribution of galaxy fields, VDM captures the inherent stochasticity of galaxy formation, enabling more reliable predictions and bridging the gap between computational efficiency and physical fidelity. 

Future work will focus on addressing the limitations observed at small scales by incorporating additional physical priors, such as baryonic effects, into the model architecture. Improving the training process through the inclusion of more diverse datasets, including simulations with varied astrophysical models, such as  SIMBA and ASTRID~\citep{simba}~\citep{astrid}, could enhance the model’s generalizability and accuracy. Furthermore, integrating the VDM framework with observational data will provide opportunities to refine its predictions and evaluate its performance in real-world applications. By continuing to refine its capabilities, the VDM can serve as a cornerstone for future cosmological analyses.

\section*{Acknowledgements}

We thank Francisco Villaescusa-Navarro for valuable discussions. TS, EG, and MR acknowledge the IT department at Michigan Technological University for their assistance in managing the computing cluster. The GPU cluster used for this work was funded by the NSF Major Research Instrumentation (MRI) Grant Award No. 221734, titled ``MRI: Acquisition of a GPU-accelerated cluster for research, training, and outreach.''
Research reported in this publication was supported in part by funding provided by the National Aeronautics and Space Administration (NASA), under award number 80NSSC20M0124, Michigan Space Grant Consortium (MSGC).

\end{document}